\documentclass[12pt]{iopart}
\usepackage{iopams}
\usepackage{graphicx}
\usepackage{epsf}

\begin{document}

\sloppy

\jl{2}

\title[Recombination of low energy electrons with U$^{28+}$]
{\bf Recombination of low energy electrons with U$^{28+}$}

\author{G. F. Gribakin and S. Sahoo}

\address{Department of Applied Mathematics and Theoretical Physics, Queens University, Belfast BT7 1NN, NI UK}
\author{V. Dzuba}
\address{School of Physics, University of New South Wales,Sydny 2052, Australia }



\begin{abstract}
We performed an extensive study of configuration mixing between the
doubly excited (doorway) states and complex multiply excited states of
U$^{28+}$ near its ionization threshold. The detailed investigation of
complex spectrum  and analysis of the statistics of eigenstate
components show that the dielectronic (doorway) states weakly mixed with
each other. However, they show  substantial mixing with the complex
multiply excited states. This situation explains the mechanism of low
energy electron recombination with U$^{28+}$. We calculated the energy
averaged capture cross sections as a sum over dielectronic doorway states and found our present calculation interprets well the experimental recombination rates in the energy range of 1 to 100 eV.  

\end{abstract}
\vspace{1cm}


\section{Introduction}
Atomic processes are of great importance from astrophysical and other various point of view. Especially the low energy recombination cross sections and rate coefficients are required for understanding the fusion and astrophysical plasmas. In particular dielectronic recombination (DR) plays a significant role in astrophysics because it is the dominant electron-ion recombination process for most ions in low density, photo-ionized and electron-ionized cosmic plasmas \cite {Arnaud:85}. Accurate rate coefficients are needed to calculate reliably the ionization balance, thermal structure and line emission of cosmic plasmas. Most importantly the low energy positron antiproton recombination provides a challenging scheme for the production of antihydrogen\cite{Holzscheiter:99}.

It is well-known that due to the presence of additional recombination
channels such as DR, the rate coefficients are found to be larger
than radiative recombination (RR) rates for many electron complex ions. In DR process the incident electron is captured in a doubly-excited state of the compound ion, which is then
stabilized by photoemission. This process originally suggested by J.~Sayers and was first considered by Massey and Bates\cite{Massey:43} in the study of ionospheric oxygen. Electron-ion recombination has been measured directly in the laboratory
since early 1980's \cite{Recomb:92}. More recently the use of heavy-ion
accelerators and electron coolers of ion storage rings has greatly advanced
the experiment \cite{Andersen:89,Kilgus:90}. Recombination rates for various
ions have been measured at electron energies from threshold to hundreds of
electron volts (eV) with a fraction-of-an-eV resolution
\cite{Schennach:94,Gao:95,Schuch:96,Uwira:96,Uwira:97,Zong:97,Mannervik:98}.
For few-electron ions the measured rates were found to be in good agreement with theoretical predictions  which included the contribution of DR resonances on top of the RR background, e.g., in He$^+$ \cite{DeWitt:94}, Li-like C$^{4+}$
\cite{Mannervik:98} and Ar$^{15+}$ \cite{Schennach:94,Zong:97}, and B-like
Ar$^{13+}$ \cite{DeWitt:96}. However, more complicated ions, e.g.,
Au$^{50+}$ \cite{Uwira:97}, U$^{28+}$\cite{Uwira:96} and
Au$^{25+}$\cite{Hoffknecht:98}, showed complicated resonance spectra and
strongly-enhanced recombination rates at low electron energies. The
Au$^{25+}$ ion has been studied in detail by Gribakin et al.\cite{Au}
and Flambaum et al.\cite{Fl:02} using statistical methods. They
suggested that the strongly enhanced low energy electron recombination
observed in this ion is mediated by complex multiply-excited states
rather than simple dielectronic resonances and the dielectronic
resonances play the role of doorways to the electron capture
process. The statistical method developed by Flambaum et al.\cite{Fl:02}
is based on the assumption of strong (chaotic) configuration
mixing. This assumption has been verified by Gribakin and Sahoo \cite{Sahoo:JPB} in a recent study. However, for U$^{28+}$, no theory so far has described the low energy DR process successfully. 

Historically the recombination rate enhancement was observed first in
U$^{28+}$\cite{Uwira:96}. This measurement has been performed in a
merged beam experiments at UNILAC accelerator in Darmstadt and at
heavy-ion storage ring TSR in Heidelberg. The experiment found rate
enhancement in the U$^{28+}$ spectrum exceeds the theoretical
calculation by at least a factor 20 in the energy range  below 10
eV. Later in 1998, they extended the experiment to high energies up to
420 eV \cite{Mitnik:98} and a comparison has been made with the
theoretical calculations which is based on distorted wave
approximations. The DR cross sections are calculated in this method are
able to explain the main resonant features in the range 80--180 eV, but
failed to identify the resonances and reproduce the rate at smaller
energies. The cross sections involving the excitations from the
5s$^{2}$5p$^{2}$ ground state configuration calculated in this method are well described by using either semirelativistic wave functions as found in AUTOSTRUCTURE codes or fully relativistic wave functions in HULLAC codes. However, the resonance structure observed at low energies i.e. below 80 eV largely remains unexplained. They finally concluded that in complex ions, particularly in U$^{28+}$ ion, what are the resonances just above the threshold and how they contribute to the low energy recombination remain a 'mystery'.   
In this paper we performed an extensive study of the excited spectra and eigenstates of U$^{28+}$ near its ionization threshold and calculated recombination rate coefficients for electron recombination with U$^{28+}$. We identified some of the resonances near the threshold those contribute significantly to the low energy recombination. The present results are found to be in good agreement with the experiment. This work develops further a statistical theory towards the full understanding of the mechanism of low energy electron recombination with U$^{28+}$ and other similar complex ions.

\section{Many-electron excitations}\label{sec:mix}
We consider the recombination of an electron with U$^{28+}$. Due to
electron correlation the slow electron can be captured in one of the
excited states of the compound ion U$^{27+}$. It has 65 electrons and
its ground state configuration belongs to
1$s^{2}$........5$s^{2}$5$p^{3}$. Figure 1 shows the spectrum of
relativistic $nlj$ orbitals obtained by relativistic Hartree-Fock
calculations. Atomic units (a.u.) are used unless otherwise stated. All
the occupied orbitals below fermi level ($\sim 31.19~ a.u.$) are obtained by a
self consistent calculation of U$^{28+}$ground state. Each of excited
orbitals above the fermi level are calculated by placing one electron on
to it in the frozen 1$s^{2}$........5$s^{2}$5$p^{2}$ (U$^{28+}$)
core. Our configuration interaction (CI) calculation shows that the
ground state of U$^{28+}$ and U$^{27+}$ ions are characterized by their
total angular momentum J $=$ 0.0 and 1.5 respectively. The difference
between their total energies $-$27741.40 a.u. and $-$27771.37 a.u.,
gives us an estimation of ionization threshold (I.T.) $=$ 29.97 a.u. (815.184 eV).

The excited states of the ion are generated by transferring one, two, three, etc., electrons from the ground state into the empty orbitals above fermi level. Since we are interested in the excited spectrum of the ion near its ionization threshold (29.97 a.u.), we consider it as a system of having only 29 electrons above the frozen Kr-like 1$s^{2}$....4$p^{6}$ core. The number of many-electron states obtained by distributing 29 electrons over 40 relativistic orbitals, 4$d_{3/2}$ through to 8$g_{9/2}$ are huge in number. It is practically impossible to perform a CI calculation for all of them. We construct the excite spectrum by using mean field approach by calculating their mean energies $E_i$, and number of many electron states $N_i$ associated with each of them:
\begin{equation}\label{eq:E}
E_{i}= E_{core}+\sum_a{\epsilon_a n_a}+\sum_{a<b}\frac{n_a(N_b-\delta_{ab})}{1+\delta_{ab}}U_{ab} ,
\end{equation}
\begin {equation}\label{eq:N}
N_i=\prod_a\frac{g_a!}{n_a! (g_a-n_a)!},
\end {equation}
where $n_a$ are the orbital occupation numbers of the relativistic
orbitals in a given configuration and
$\sum_a{n_a}=n$. $\epsilon_a=<a|H_{core}|a>$ is the s


ingle-particle energy of the orbital $a$ in the field of the core, $g_a=2j_a+1$, and $U_{ab}$ are the average Coulomb matrix elements for the electrons in orbitals $a$ and $b$ ( direct minus exchange):

\begin{equation}\label{eq:U}
U_{ab}=\frac{g_a}{g_a-\delta_{ab}}\left[R_{abab}^{(0)}-\sum_{\lambda}\delta_p R_{abba}^{(\lambda)}\left\{ {j_a \atop \frac{1}{2}}{\j_b \atop-\frac{1}{2}}{\lambda \atop \ 0}\right\}^{2}\right]
\end{equation}

$R_{abba}^{(\lambda)}$ is the two-body radial Coulomb integral of $\lambda$ multipole, and $\delta_p=1$ when $l_a +l_b +\lambda$ is even and 0 otherwise. Using Eqs. (1)-(3) we obtained about 353 configurations within $\pm$ 1 a.u. of ionization threshold. They comprise a total of $1.9\times 10^5$ many electron states.
The single-particle spectrum of U$^{28+}$ does not show large gaps. Owing to the ``gapless''single-particle spectrum, the density increases rapidly as a function of energy, as described by the Fermi-gas-model ansatz
\cite{Bohr:69}
\begin{equation}\label{eq:rho}
\rho (E)=AE^{-\nu }\exp (a \sqrt{E}),
\end{equation}
with $A$ = 0.0885, $\nu$ = 2.33, and $a$ = 3.99 a.u.\cite{comment3},
where $E$ is the energy above the ground state in atomic units. 
Figure 2 shows the level density calculated by averaging with a Gaussian with 1
a.u. variance. This figure depicts the level densities for both odd and
even parity configurations. Also included in the figure are only the doubly
excited configurations of even and odd parity found within $\pm$ 1
a.u. of ionization threshold. It is found that the level density of the
odd configurations near its ionization threshold is of the order is of
$1\times 10^5$ and 

\begin{figure}[h]
\epsfxsize=10cm
\centering\leavevmode\epsfbox{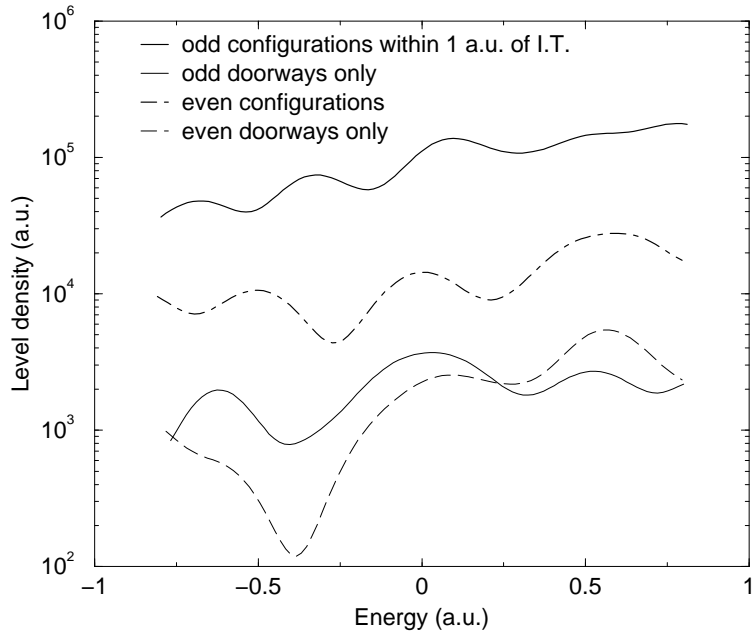}
\vspace{8pt}
\caption{Level densities in U$^{28+}$ within $\pm 1$ a.u. of Ionization threshold. Thick solid line: odd configurations, Thin solid line: odd dielectronic configurations, Dash-dot line: even configurations and dashed line: even dieclectronic configurations.}
\end{figure}
those of
even configurations it is of the order of $1.3\times 10^4$. This
provides an evidence that this system is characterized by huge level
density. This system may be compared with Au$^{25+}$, in which the level
density near its ionization threshold is about $1\times
10^7$\cite{Au}. This shows that  U$^{28+}$ ion is less
complicated than Au$^{25+}$ from the view point of dense excited spectrum. Apart from this one can estimate the mean level spacing from the distribution of total angular momentum J as shown in Figure 3. 

Ionic eigenstates are characterized by their total angular momentum and
parity $J^\pi $, and are $2J+1$ times degenerate. Therefore the total level
density can be broken into a sum of partial level densities:
$\rho (E)=\sum _{J^\pi }(2J+1)\rho _{J^\pi }(E)$. The  excitation spectrum of
U$^{28+}$ near the ionization threshold, $E=I\approx 29.97$ a.u., 
contains many $J$ ranging from $\frac{1}{2}$ to $\frac{25}{2}$.
Their distribution is in agreement with statistical theory
\cite{Bohr:69,Bauche:87}, which predicts that at a given energy $\rho _{J^\pi }$
are proportional to the function
\begin{equation}\label{eq:f_J}
f(J)= \frac{2(2J+1)}{(2J_m+1)^2}\exp \left[ -\frac{(2J+1)^2}
{2(2J_m+1)^2}\right] ,
\end{equation}
where $J_m$ is the most abundant $J$ value. Numerically for U$^{27+}$ we find
$J_m \approx \frac{7}{2}$.
\begin{figure}[h]
\epsfxsize=10cm
\centering\leavevmode\epsfbox{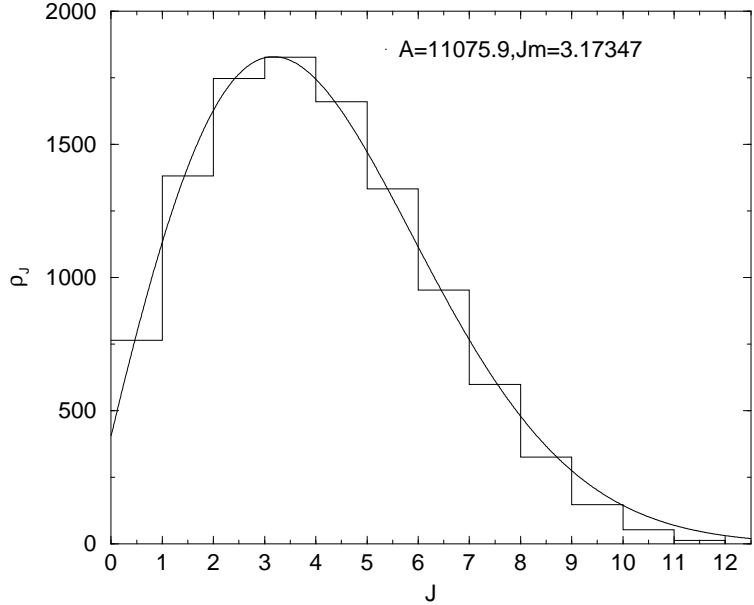}
\caption{Densities of states with different J near ionization threshold
of  U$^{28+}$. Histograms show their distribution and the solid line is
non linear fitting obtained from Eq.(5) which predicts the most abundant
value of $J\sim3.2$ and the proportionality constant $\pm$ 11075.9.} 
\end{figure}

Using Eq. (\ref{eq:f_J}) one can estimate the partial densities by
$\rho _{J}=\rho _{J^+}+\rho _{J^-}\simeq f(J)\rho / \langle 2J+1\rangle $,
where $\langle 2J+1\rangle $ is an average over $f(J)$. For the most abundant
angular momenta $J\sim J_m$, and assuming $\rho _{J^+}\approx \rho _{J^-}$,
we have $\rho _{J^\pi }(E)=A_{J^\pi }E^{-\nu }\exp (a \sqrt{E})$.
Near the ionization threshold this gives
$\rho _{J^\pi }\approx 1.8\times 10^3$ a.u. (Fig. 3), which means that the
spacing between the multiply-excited states with a given $J^\pi $ is 
small: $D=1/\rho _{J^\pi }\sim 15$ meV. Whereas in Au$^{25+}$ it is about 1 meV, i.e. 15 times larger than that found in Au$^{25+}$. This situation explains why individual resonances appear in recombination rates of U$^{28+}$  but the same is not observed in Au$^{25+}$ experimentally even at an energy resolution of 0.1 eV. However, the large density of
multiply-excited states is only a ``kinematic'' reason behind the
experimental observation. To explain the fact that the electron can actually be captured into these states, we need to analyze the dynamics
of electron capture and show that the residual Coulomb interaction between the
electrons (i.e. that beyond the mean field) makes for an efficient capture and
accounts for the observed enhanced recombination rate.
\section{Configuration mixing}
Taking into account the fact that the rasidual Coulomb interaction is the key problem in many-electron
processes, we construct the basis of many-electron
states $\Phi _k $ from single-particle (e.g., Dirac-Fock) orbitals, and
solve the eigenvalue problem for the Hamiltonian matrix
$H_{ik}=\langle \Phi _i|\hat H|\Phi _k\rangle $, which yields the
eigenvalues $E_\nu $ and eigenstates $|\Psi _\nu \rangle =\sum _kC^{(\nu )}_k
|\Phi _k \rangle $ of the system (configuration interaction method).
We performed two sets of model calculations. One includs all the
configurations within $\pm 1$ a.u. of ionization threshold of U$^{28+}$,
which produces 2516 states for $J^{\pi}$=$\frac{7}{2}^{-}$
sequence. Similarly the second calculation includes all the dielectronic
(doorway) states within $\pm 1$ a.u. of ionization threshold and produces 108
states for  $J^{\pi}$=$\frac{7}{2}^{-}$ sequence. This shows that the
number of states associated with the dielectronic configurations for a
given $J^{\pi}$ are not large in number. As a result we performed the
full CI calculations for Hamiltonian matrix of size 108 and 2516
respectively and obtained the eigen values and eigenstate components. To
study the mixing between the doubly excited states we analyze the
eigenstate components by calculating the weight of a given doorway
configuration shown in Figure 4(a)-(e). The weight ($w$) of a doorway
state can be calculated as $\Sigma_{k=1,Nc}|C^{\nu}_{k}|^{2}$, where
$N_{c}$ is the number of states in each configuration. It has been
shown in Ref\cite{Sahoo:JPB} that when there is strong and uniform
configuration mixing the
weights significantly reduce from 1, but in the present case the weights
go down from 1 but not significantly. So one can say that these doubly
excited (doorways) states weakly mix with each other. When they are
included in the large calculation (2516 states), their weights
significantly go down from 1 as shown in Figure 4(f)-(j). This gives a
signature of strong mixing i.e., the doorways show a significant mixing with multiply excited states. It has also been
observed that the multiply excited configurations mix with each other
quite comfortably. This mixing is mainly  responsible for an enhancement of  recombination rates
over RR. In the recombination cross sections, the doorways which do not mix completely with either multiply excited states or with other doorways (in a sense they remain 'isolated') appear in the form of narrow peaks (resonances) which explains the experiment as well as the theory\cite{Mitnik:98}. 

On the other hand, when the level density is high and the two-body interaction
is sufficiently strong the system is driven into a regime of {\em many-body
quantum chaos}, where the effect of configuration mixing can be described
statistically in the case of Au$^{25+}$\cite{Au}. This regime is characterized by the following
\cite{Zel,Ce}. (i) Each eigenstate contains a large number $N$ of
{\em principal} components $C^{(\nu )}_k\sim 1/\sqrt{N}$, corresponding to
the basis states $\Phi _k $ which are strongly mixed together. (ii) Owing to
the strong mixing, the only good quantum numbers that can be used to classify
the eigenstates, are the exactly conserved total angular momentum and parity
$J^\pi $ and the energy. (iii) The degree of mixing in this regime is in some
sense complete, i.e. all basis states that can be mixed (within a
certain energy rage, see below) are mixed. The notion of configurations based
on the single-particle orbitals becomes largely irrelevant for the purpose of
classifying the eigenstates. Each eigenstate contains substantial
contributions of a few nearby configurations. As mentioned above, in the present case we
found that there is a weak configuration mixing  between the
dielectronic doorway configurations and complicated multiply excited
states, and the multiply excited states show a substantially strong
configuration mixing with each other. But the degree of mixing is not
sufficiently strong to drive the system into chaotic regime.

\begin{figure}[h]
\epsfxsize=10cm
\centering\leavevmode\epsfbox{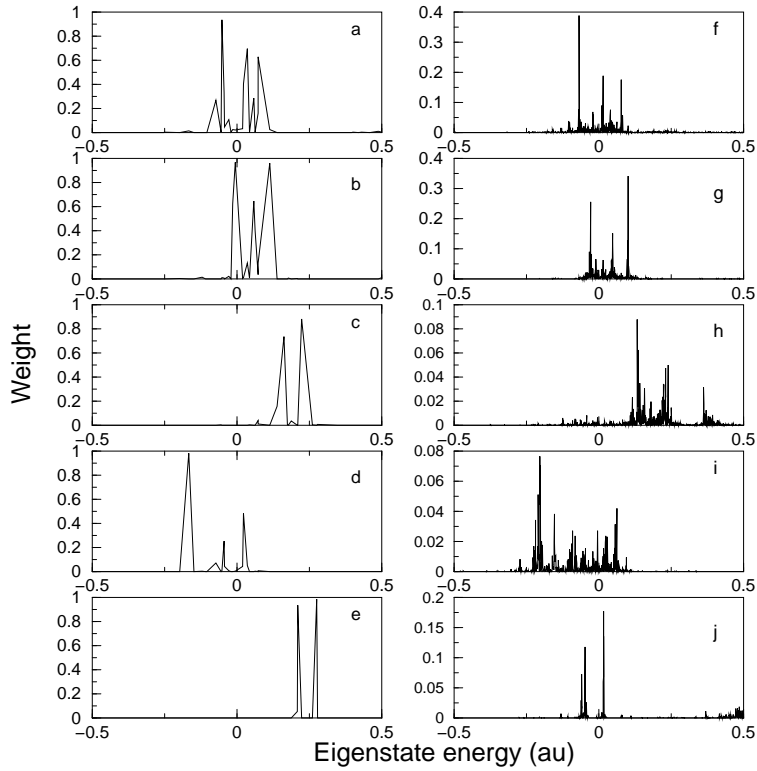}
\caption{Weights of odd dielectronic doorway states obtained from two different calculations. Figure (a)-(e) are the weights of doorways obtained from a calculation involving 108 states. Figure (f)-(j) are the weights of the same doorways obtained from a calculation involving 2516 states}.

{(a):$4d_{3/2}^{4}$$4d_{5/2}^{6}$$4f_{5/2}^{6}$$4f_{7/2}^{\bf7}$$5s_{1/2}^{2}$$5p_{1/2}^{2}$$5p_{3/2}^{1}$$7f_{5/2}^{1}$($N_{c}$=$4$)
(b):$4d_{3/2}^{4}$$4d_{5/2}^{6}$$4f_{5/2}^{6}$$4f_{7/2}^{\bf7}$$5s_{1/2}^{2}$$5p_{1/2}^{2}$$5p_{3/2}^{1}$$7f_{7/2}^{1}$($N_{c}$=$4$)
(c):$4d_{3/2}^{4}$$4d_{5/2}^{6}$$4f_{5/2}^{6}$$4f_{7/2}^{8}$$5s_{1/2}^{2}$$5p_{1/2}^{\bf1}$$5g_{9/2}^{1}$$6p_{3/2}^{1}$($N_{c}$=$2$)
(d):$4d_{3/2}^{4}$$4d_{5/2}^{6}$$4f_{5/2}^{6}$$4f_{7/2}^{8}$$5s_{1/2}^{\bf1}$$5p_{1/2}^{2}$$5d_{5/2}^{1}$$7f_{5/2}^{1}$($N_{c}$=$2$)
(e):$4d_{3/2}^{4}$$4d_{5/2}^{6}$$4f_{5/2}^{\bf5}$$4f_{7/2}^{8}$$5s_{1/2}^{2}$$5p_{1/2}^{2}$$5f_{5/2}^{1}$$6p_{1/2}^{1}$($N_{c}$=$2$)}

{$N_{c}$ is the number of states associated with each of the configurations}.
\end{figure}

This feature can be studied from the inverse
participation ratio (IPR): $\Sigma_{j=1,Nc}|C^{\nu}_{k}|^{-4}$. 
\begin{figure}[h]
\epsfxsize=10cm
\centering\leavevmode\epsfbox{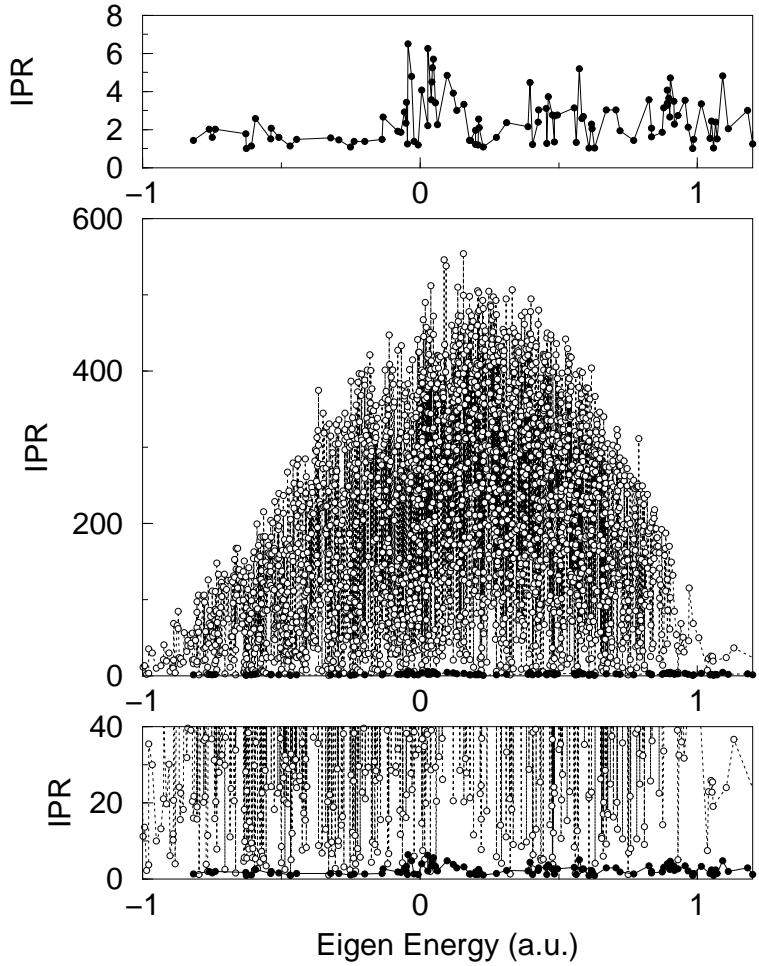}
\caption{Inverse participation ratio (IPR). Top figure: for doorways
having 108 states, the middle figure: for all configurations having
2516 states and the bottom one is same as middle figure but in
an extended scale.}
\end{figure}
Figure 5
shows the inverse participation ratio for the two sets of model
calculations. The top figure depicts the inverse participation ratio of
the doorways only and the middle one for the large (2516 states)
calculation that includes all the configurations (both dielectronic and
multiply excited states). We found most of the doorways are associated
with number of states either 4 or 2 or even less. The top figure shows
that the inverse participation ratio  lies flat between 2 and 4 except
at the energy range where the eigen energies are close to the ionization
threshold. This indicates that a few number of the doorways having
energies close to
the threshold participate in mixing. However the middle figure shows a
picture of strong but non uniform mixing which involves a lots of multiply excited
states. When we included the IPR of doorways with the IPR obtained from
large calculation as shown in the bottom figure, it 
lies well below as indicated by solid circles. It also provides us with
an information that some of the doorways take part in mixing with multiply excited
configurations and other doorways are weakly mixed or even remain isolated.   

This strong mixing takes place in a certain energy range $|E_k-E_\nu |\lesssim
\Gamma _{\rm spr}$, where $E_k\equiv H_{kk}$ is the mean energy of the basis
state and $\Gamma _{\rm spr}$ is the so-called {\em spreading width}.
More precisely, the mean-squared value of $C^{(\nu )}_k$
as a function of $E_k-E_\nu $, can be described by a Breit-Wigner (BW) formula
\begin{equation}\label{eq:C^2}
\overline {\left| C_k^{(\nu)}\right|^2}=N^{-1}\frac{\Gamma _{\rm spr}^2/4}
{(E_k-E_\nu )^2+\Gamma _{\rm spr}^2/4},
\end{equation}
with $N=\pi \Gamma _{\rm spr}/2D$ fixed by normalization
$\sum _k\left| C_k^{(\nu)}\right|^2\simeq
\int \overline {\left| C_k^{(\nu)}\right|^2} dE_k/D=1$.

\begin{figure}[h]
\epsfxsize=10cm
\centering\leavevmode\epsfbox{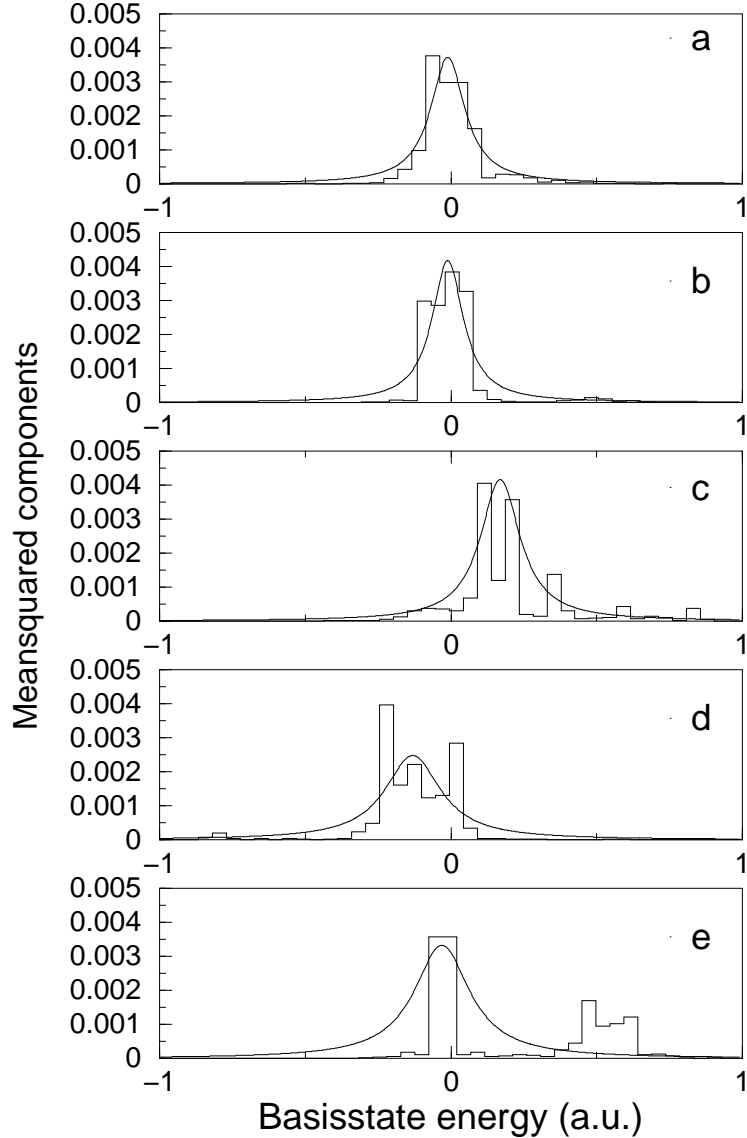}
\caption{Spreading of doorway configurations in complicated multiply
excited states obtained from 2516 x 2516 calculations for
$J^{\pi}=3.5^{-}$ sequence. Each of the doorway configuration spreads as (a) 0.1433 a.u., (b) 0.1310 a.u., (c) 0.1661 a.u., (d) 0.2321 a.u. and (e) 0.2310 a.u.. These doorways are same as indicated in Fig4.}

\end{figure}

 The mean-squared components are obtained by averaging over the
basis states associated with each of the doorway configurations and are
plotted as a function of eigen energies.  We obtained $\Gamma _{\rm
spr}$ of doorways (close to the
ionization threshold) from the BW fit which is shown in Figure
6(a)-(e). From the BW fit we observed
that the $\Gamma _{\rm spr}$ is not constant and it varies from 0.1
a.u. to 0.2 a.u. in these doorways. There may be doorways which even
show much less spreading. Roughly one can say the mixing takes place
within 0.1 a.u.. So we calculated the recombination rates with two
different values of spreading width which can be found in section IV. It
may be pointed out that though the single BW fit is not as accurate as it should be, still it gives an estimation of the important quantity i.e. $\Gamma_{\rm spr}$. It is worth mentioning that the spreading width found in Au$^{25+}$ is about 0.5 a.u.\cite{Au} and these value does not change if one performs calculations by adding more configurations\cite{Sahoo:JPB}.

\section{Recombination}\label{sec:rec}
\subsection{Theory}
For low-energy electrons the contribution of the autoionising
states (resonances) to the recombination cross section is given by 
(see, e.g., Ref. \cite{Landau})
\begin{equation}\label{eq:sigma_res}
\sigma _r=\frac{\pi }{k^2}\sum _\nu \frac{2J+1}{2(2J_i+1)}\,
\frac{\Gamma _\nu ^{(r)}\Gamma _\nu ^{(a)}}{(\varepsilon -\varepsilon _\nu )^2
+\Gamma _\nu ^2/4},
\end{equation}
where $\varepsilon =k^2/2$ is the electron energy, $J_i$ is the angular
momentum of the initial (ground) target state, $J$ are the angular momenta
of the resonances, $\varepsilon _\nu =E_\nu -I$ is the position of the
$\nu $th resonance relative to the ionization threshold of the compound
(final-state) ion, and $\Gamma _\nu ^{(a)}$, $\Gamma _\nu ^{(r)}$,
and $\Gamma _\nu =\Gamma _\nu ^{(r)}+\Gamma _\nu ^{(a)}$ are its
autoionisation, radiative, and total widths, respectively
\cite{comment4}. When the resonance spectrum is dense, $\sigma _r$
can be averaged over an energy interval $\Delta \varepsilon$ which 
contains many resonances,
$D\ll \Delta \varepsilon \ll \varepsilon $, yielding
\begin{equation}\label{eq:sigres_av}
\bar \sigma _r=\frac{2\pi ^2}{k^2}\sum _{J^\pi }
\frac{2J +1}{2(2J_i+1)D}\left\langle \frac{\Gamma _\nu ^{(r)}
\Gamma _\nu ^{(a)}}{\Gamma _\nu ^{(r)}+\Gamma _\nu ^{(a)}} \right\rangle ,
\end{equation}
where $\langle \dots \rangle $ means averaging. If the fluorescence yield,
$\omega _f\equiv \Gamma _\nu ^{(r)}/(\Gamma _\nu ^{(r)}+\Gamma _\nu ^{(a)})$,
fluctuates weakly from resonance to resonance (see below), one can
write $\bar \sigma _r=\bar \sigma _c\omega _f$, where
\begin{equation}\label{eq:sig_cap}
\bar \sigma _c=\frac{\pi ^2}{k^2}\sum _{J^\pi } \frac{(2J +1)
\Gamma ^{(a)}}{(2J_i+1)D}
\end{equation}
is the energy-averaged capture cross section, and $\Gamma ^{(a)}$ is the
average autoionisation width.

In a situation when there is a strong configuration mixing between the
dielectronic doorway states and multiply excited states, the capture
cross sections can be obtained as a  sum over the single-electron excited states
$\alpha ,~\beta $ and hole states $\gamma $, as well as the partial waves
$lj$ of the continuous-spectrum electron $\varepsilon $. As a result, we have
\begin{eqnarray}\label{eq:sig_cap2}
\bar \sigma _c&=&\frac{\pi ^2}{k^2}\sum _{\alpha \beta \gamma ,lj}
\frac{\Gamma _{\rm spr}}{(\varepsilon -\varepsilon _\alpha -
\varepsilon _\beta +\varepsilon _\gamma )^2 +\Gamma _{\rm spr}^2/4}
\sum _\lambda \frac{\langle \alpha ,\beta \| V_\lambda \| \gamma ,
\varepsilon lj \rangle }{2\lambda +1}\nonumber \\
&\times &\Biggl[ \langle \alpha ,\beta \| V _\lambda \| \gamma ,
\varepsilon lj\rangle -(2\lambda +1) \sum _{\lambda '}
(-1)^{\lambda +\lambda '+1}\left\{ {\lambda \atop \lambda '}{j_\alpha 
\atop j_\beta }{j \atop j_\gamma }\right\} \langle \alpha ,\beta \| V
_{\lambda '}\| \varepsilon lj,\gamma \rangle \Biggr] ,
\end{eqnarray}
where $\varepsilon _\alpha $, $\varepsilon _\beta $ and $\varepsilon _\gamma $
are the orbital energies, the two terms in square brackets represent the
direct and exchange contributions, and
$\langle \alpha ,\beta \| V_\lambda \| \gamma ,
\varepsilon lj \rangle $ is the reduced Coulomb matrix element (see Ref.\cite{Fl:02}).

It is assumed that the energies of dielectronic doorway states relative
to the threshold is given by $\varepsilon _\alpha $ + $\varepsilon
_\beta $ - $\varepsilon _\gamma$. 
A more accurate value can be obtained by using mean field energies
 (configuration energies) of
doorway configurations in Eq.(10).  The effect of using these two
different energies can be found in next subsection. We have also shown 
the the qualitative difference between the results obtained by using two
different values of $\Gamma_{spr}$.  Because $\Gamma_{spr}$ is well defined in
the case of a strong and chaotic configuration mixing. However, it can
not be 
properly defined if the mixing of the configurations  is weak and non-uniform.    

Equation (\ref{eq:sig_cap2}) is directly applicable to targets with
closed-shell ground states. If the target ground state contains partially
occupied orbitals, a factor
\begin{equation}\label{eq:occup}
\frac{n_\gamma }{2j_\gamma +1}\left(1-\frac{n_\alpha }{2j_\alpha 
+1}\right)\left(1-\frac{n_\beta }{2j_\beta  +1}\right),
\end{equation}
where $n_\alpha $, $n_\beta $, and $n_\gamma $ are the orbital occupation
numbers in the ground state $\Phi _i$, must be introduced on the right-hand
side of Eq. (\ref{eq:sig_cap2}). Steps similar to those that lead to
Eq. (\ref{eq:sig_cap2}) were used to obtain mean-squared matrix elements of
operators between chaotic many-body states \cite{Ce,Flambaum:93}.

The chaotic nature of the multiply-excited states $\Psi _\nu $ can also be
employed to estimate their radiative widths $\Gamma _\nu ^{(r)}$. 
Electron-photon interaction is described by a single-particle dipole operator
$\hat d$. Any excited electron in $\Psi _\nu $ may emit a photon, thus leading
to radiative stabilization of this state. The total photo-emission rate
$\Gamma _\nu ^{(r)}$ can be estimated as a weighted sum of the single-particle
rates,
\begin{equation}\label{eq:Gamma_r}
\Gamma _\nu ^{(r)}\simeq \sum _{\alpha ,\beta }
\frac{4\omega _{\beta \alpha }^3} {3c^3}
|\langle \alpha \|\hat d\|\beta \rangle |^2
\left\langle \frac{n_\beta }{2j_\beta +1}\left( 1-\frac{n_\alpha }
{2j_\alpha +1}\right) \right\rangle _\nu ,
\end{equation}
where $\omega _{\beta \alpha }=\varepsilon _\beta -\varepsilon _\alpha >0$,
$\langle \alpha \|\hat d\|\beta \rangle $ is the reduced dipole operator
between the orbitals $\alpha $ and $\beta $, and
$\langle \dots \rangle _\nu $ is the
mean occupation number factor. Since $\Psi _\nu $ have large numbers
of principal components $N$, their radiative widths display small $1/\sqrt{N}$
fluctuations. This can also be seen if one recalls that a chaotic
multiply-excited state is coupled by photo-emission to many lower-lying states,
and the total radiative width is the sum of a large number of (strongly
fluctuating) partial widths. A similar effect is known in compound nucleus
resonances in low-energy neutron scattering \cite{Bohr:69}.

There is a certain similarity between Eqs. (\ref{eq:sig_cap2}) and
(\ref{eq:Gamma_r}) and those for autoionisation and radiative rates obtained
in a so-called configuration-average approximation \cite{Griffin:85}.
In both cases the answers involve squares or products of two-body Coulomb
matrix elements [see the direct and exchange terms in Eq. (\ref{eq:sig_cap2})],
or single-particle dipole amplitudes [Eq. (\ref{eq:Gamma_r})]. However, there
are a number of important differences between the present results and
the configuration-average approximation. The latter considers dielectronic
recombination and introduces averaging over configurations as a means of
simplifying the calculation. The DR cross section is averaged over an
arbitrary energy interval $\Delta \varepsilon $, and only the configurations
within this energy range contribute to the average. Effects of configuration
mixing as well as level mixing within a configuration are neglected.

It is important to compare the radiative and autoionisation widths of
chaotic multiply-excited states. Equation (\ref{eq:Gamma_r}) shows that
$\Gamma ^{(r)}$ is comparable to the single-particle radiative widths.
On the other hand, the autoionisation width $\Gamma ^{(a)}$, is suppressed by a factor
$\left| C_k^{(\nu )}\right| ^2\sim N^{-1}$ relative to that of a typical
dielectronic resonance.  Therefore, in systems
with dense spectra of chaotic multiply-excited states the autoionisation
widths are small. Physically this happens because the coupling strength
of a two-electron doorway state to the continuum is shared between many
complicated multiply-excited eigenstates.
As a result, the radiative width may dominate the total width of the
resonances, $\Gamma ^{(r)}\gg \Gamma ^{(a)}$, making their
fluorescence yield close to unity. However, in the case of U$^{28+}$, $\Gamma ^{(r)}\sim \Gamma ^{(a)}$.  Our numerical results for the
recombination of U$^{28+}$ presented in Sec. \ref{sec:num}, confirm this
scenario.

The resonance recombination cross section should be compared with the 
direct radiative recombination cross section
\begin{equation}\label{eq:sigmad}
\sigma _d= \frac{32\pi }{3\sqrt{3}c^3}\,\frac{Z_i^2}
{k^2} \ln \left( \frac{Z_i}{n_0k}\right) ,
\end{equation}
obtained from the Kramers formula by summing over the principal
quantum number of the final state \cite{Au}. Here $Z_i$ is the ionic charge
($Z_i=28$ for U$^{28+}$), and $n_0$ is the principal quantum number of
the lowest unoccupied ionic orbital ($n_0=5$). Note that the direct 
and energy-averaged resonance recombination cross sections of
Eqs. (\ref{eq:sigmad})
and  (\ref{eq:sigres_av}) have similar energy dependences.

\subsection{Numerical results}\label{sec:num}

Numerical calculations of the cross section from Eqs. (\ref{eq:sig_cap2})
and (\ref{eq:occup}) involve summation over the orbitals shown in
Fig. 1 with electron partial waves up to $i_{11/2}$. The results of the calculations for the
recombination rates are displayed in Figure 7. We calculated  the
recombination rates using two different energies, i. e., configuration
energies and relativistic energies. It is found that both the
calculations are in good agreement. However, the magnitude of
recombination rates obtained from both the present calculations lie
above the experimental rates. This is due to the fact that we used
$\omega_{f}$ = 1 in the present calculations similar to that in
Au$^{25+}$. It is clear that in the present case $\omega_{f}$ is less
than 1 and remains constant throughout the energy range considered. The
exchange contributions are found to be about  200 times smaller than the
direct one so we did not include these when calculated the final rates. The
radiative rates are found to be smaller in magnitude in comparison to DR
rates and have almost negligible interference with the resonances. It may
be pointed out that the present calculation with
spreading width  $\Gamma_{spr}$ $\sim 0.15$ a.u, does not show enough
resonance structures as have been observed in the experiment. Hence, we
performed another calculation with  $\Gamma_{spr}$ $\sim
0.05 $ a.u. and compared with the results obtained using spreading with
0.15 a.u. shown in Figure 8. It is clear that the results do not show
any change in magnitude. However, the calculation with  $\Gamma_{spr}$
$\sim 0.05$ interprets the resonance peaks
very well which are in reasonable agreement with the experiment, though
the position of peaks are different from the experiment. This is because the
present energies of the doorway states are approximate and a few eV
relative error is expected in this approach. The presence of narrow peaks in
rate coefficients can be interpreted as: the dielectronic states which
play the role doorway to the electron capture process weakly mix with
each other as has been discussed in previous section and appear as
single peaks. Because in statistical calculations one peak corresponds
to one doorway state.  It may be recalled that the
distorted wave calculation\cite{Mitnik:98}, predicts the recombination
rates in agreement with the experiment above 80 eV and below this energy
the results are smaller in magnitude.
\begin{figure}[h]
\epsfxsize=10cm
\centering\leavevmode\epsfbox{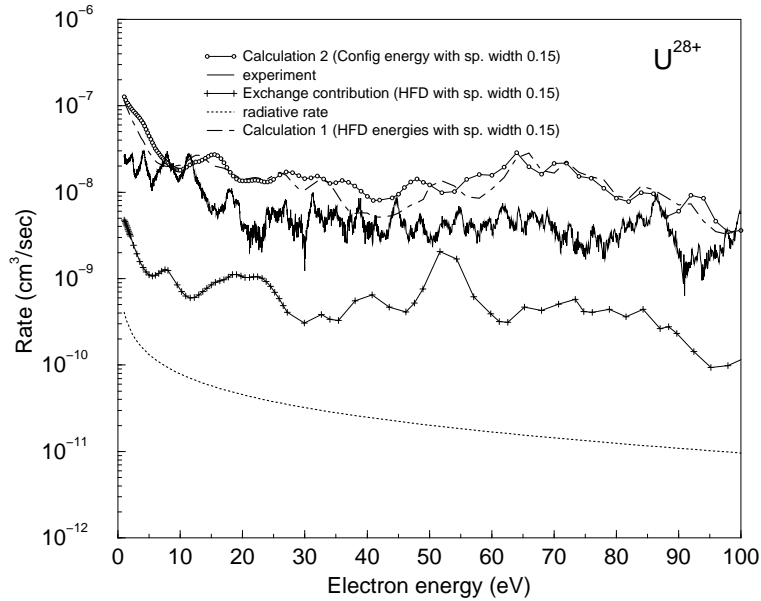}
\caption{Electron recombination rate on U$^{28+}$. Open circles
connected by solid line: Calculation 2 (using configuration energy and
$\Gamma_{spread}=0.15$ a.u.), Dot-dashed line: Calculation 1 (using HFD
energy and $\Gamma_{spread}=0.15$ a.u.), Solid lines : Experimental
data, Solid line connected by plus sign : Exchange contribution (using
HFD energy and $\Gamma_{spread}=0.15$ a.u.), Dotted line: Radiative rate.}
\end{figure}
\begin{table}
\caption{Electron orbitals which give the leading contribution to the low-energy (around 1 eV) electron recombination on U$^{28+}$.}
\label{tab:door}
\begin{tabular}{ccccccc}

\multicolumn{4}{c}{Orbitals[1]} & Direct &
$\Delta E$[3] & $\Delta E$[4]\\
\cline{1-4}
$\alpha $ & $\beta $ & $\gamma $ & $\varepsilon lj$ &
contribution[2] & (a.u.) & (a.u)\\
\hline
$6p_{1/2}$ & $5f_{7/2}$ & $4f_{7/2}$ & $p_{1/2}$ & 0.0117 & $-$0.132 & $-$0.114\\
$6p_{1/2}$ & $5f_{5/2}$ & $4f_{5/2}$ & $p_{1/2}$ & 0.0145 & 0.136 & 0.146 \\

$5p_{3/2}$ & $7f_{7/2}$ & $4f_{7/2}$ & $p_{3/2}$ & 0.0292 & 0.085 & 0.036 \\

$8s_{1/2}$ & $5d_{5/2}$ & $5s_{1/2}$ & $d_{3/2}$ & 0.0090 & 0.065 & 0.064 \\

$5d_{3/2}$ & $6f_{7/2}$ & $4f_{7/2}$ & $d_{3/2}$ & 0.0707 & $-$0.033 & 0.056 \\

$5d_{3/2}$ & $6f_{5/2}$ & $4f_{5/2}$ & $d_{3/2}$ & 0.0019 & 0.341 &0.438 \\

$8s_{1/2}$ & $5d_{5/2}$ & $5s_{1/2}$ & $d_{5/2}$ & 0.0134 & 0.065 & 0.064 \\

$5d_{5/2}$ & $6f_{7/2}$ & $4f_{7/2}$ & $d_{5/2}$ & 0.0019 & 0.627 & 0.602 \\

$6p_{1/2}$ & $6f_{5/2}$ & $5p_{1/2}$ & $f_{5/2}$ & 0.0540 & $-$0.226 & $-$0.004 \\
$8p_{1/2}$ & $5d_{3/2}$ & $5s_{1/2}$ & $f_{5/2}$ & 0.0024 & $-$0.204 & $-$0.110\\

$7f_{5/2}$ & $5d_{5/2}$ & $5s_{1/2}$ & $f_{5/2}$ & 0.0015 & $-$0.089 &$-$0.096 \\

$8g_{7/2}$ & $5d_{3/2}$ & $5p_{1/2}$ & $f_{5/2}$ & 0.0083 & $-$0.087 & 0.133 \\

$8p_{3/2}$ & $5d_{3/2}$ & $5s_{1/2}$ & $f_{7/2}$ & 0.0018 & 0.155 & 0.264 \\

$6f_{7/2}$ & $6p_{1/2}$ & $5p_{1/2}$ & $f_{7/2}$ & 0.0758 & $-$0.158 & 0.073 \\

$7f_{7/2}$ & $5d_{5/2}$ & $5s_{1/2}$ & $f_{7/2}$ & 0.0033 & $-$0.052 &$-$0.060 \\

$5g_{9/2}$ & $6d_{3/2}$ & $5p_{1/2}$ & $f_{7/2}$ & 0.0011 & $-$0.159 &$-$0.119 \\

$6d_{3/2}$ & $5g_{7/2}$ & $5p_{1/2}$ & $h_{9/2}$ & 0.0151 & $-$0.196 &$-$0.148 \\

$6d_{5/2}$ & $5g_{7/2}$ & $5p_{1/2}$ & $h_{9/2}$ & 0.0028 & 0.040 & 0.129 \\

$5f_{5/2}$ & $6p_{1/2}$ & $4f_{5/2}$ & $h_{9/2}$ & 0.0014 & 0.136 & 0.146 \\

$7f_{5/2}$ & $5d_{5/2}$ & $5s_{1/2}$ & $h_{9/2}$ & 0.0094 & $-$0.090 &$-$0.096
\\

$7f_{5/2}$ & $5p_{3/2}$ & $4f_{7/2}$ & $h_{9/2}$ & 0.0049 & 0.048 & $-$0.003 \\

$5g_{7/2}$ & $6p_{3/2}$ & $5s_{1/2}$ & $h_{9/2}$ & 0.0017 & 0.447 & 0.538 \\

$8g_{7/2}$ & $5d_{3/2}$ & $5p_{1/2}$ & $h_{9/2}$ & 0.0566 & $-$0.087 & 0.133 \\

$5d_{5/2}$ & $7f_{5/2}$ & $5s_{1/2}$ & $h_{11/2}$ & 0.0012 & $-$0.089 & $-$0.096\\

$6d_{5/2}$ & $5g_{7/2}$ & $5p_{1/2}$ & $h_{11/2}$ & 0.0131 & 0.040 & 0.013 \\

$7f_{7/2}$ & $5d_{5/2}$ & $5s_{1/2}$ & $h_{11/2}$ & 0.0193 & $-$0.052 & $-$0.060 \\

$7f_{7/2}$ & $5p_{3/2}$ & $4f_{7/2}$ & $h_{11/2}$ & 0.0084 & 0.085 & 0.036 \\

$5g_{9/2}$ & $6d_{3/2}$ & $5p_{1/2}$ & $h_{11/2}$ & 0.0242 & $-$0.159 & $-$0.118 \\

$5g_{9/2}$ & $6p_{3/2}$ & $5s_{1/2}$ & $h_{11/2}$ & 0.0018 & 0.485 & 0.573 \\

$5g_{9/2}$ & $6d_{5/2}$ & $5p_{1/2}$ & $h_{11/2}$ & 0.0028 & 0.078 & 0.147 \\

$6g_{9/2}$ & $5f_{5/2}$ & $5p_{1/2}$ & $i_{11/2}$ & 0.0053 & 0.084 & 0.157 \\

\end{tabular}
[1]{$\alpha $ and $\beta $ are the excited electron
orbitals, and $\gamma $ is the ground-state hole of the dielectronic
doorway state; $\varepsilon lj$ is the partial wave of the incident electron.}
[2]{Direct term contributions to the dimensionless sum in
Eq.~(\ref{eq:sig_cap2}), $\bar \sigma _c k^2/\pi ^2$, using spreading width = 0.15 a.u. and configuration energy , with magnitudes greater than $5\times 10^{-3}$.}
[3]{$\Delta E=\varepsilon _\alpha +\varepsilon _\beta -
\varepsilon _\gamma $ is the mean-field energy of the doorway state relative
to the threshold.}
[4]{$\Delta E$ is the configuration energy relative to the
threshold.}
\end{table}
  
Table \ref{tab:door} lists the most important dielectronic doorway
contributions to the dimensionless sum in Eq.~(\ref{eq:sig_cap2}), which
also determines the ratio of the autoionisation width to the spacing
between the resonances. In total they account for about
two thirds of the total cross section. Although these transitions 
have been selected according to the size
of their contribution, their energies from two different calculations are close to
the threshold, as seen in the last two columns in Table \ref{tab:door}. Indeed,
the spreading of configurations discussed in Sec. \ref{sec:mix} allows
configurations near the threshold, $|\Delta E|\lesssim \Gamma _{\rm spr}$,
to contribute. On the other hand, the contribution of configurations lying
far away from threshold, $|\Delta E|\gg \Gamma _{\rm spr}$, is suppressed. It may be pointed out that the theory of Mitnik et. al.\cite{Mitnik:98} considers the excitations from 5$s$ orbital only. However, the present calculations show that excitations involving 4$f$ orbitals are  more important especially at threshold since they contribute significantly to the recombination cross sections.
As expected, the energy dependences of the resonance and direct recombination
rates are very close, although the latter is about 200 times smaller.
There is a good overall agreement between the resonance rate and experimental
data at electron energies between 1 eV and 100 eV.
\begin{figure}[h]
\epsfxsize=10cm
\centering\leavevmode\epsfbox{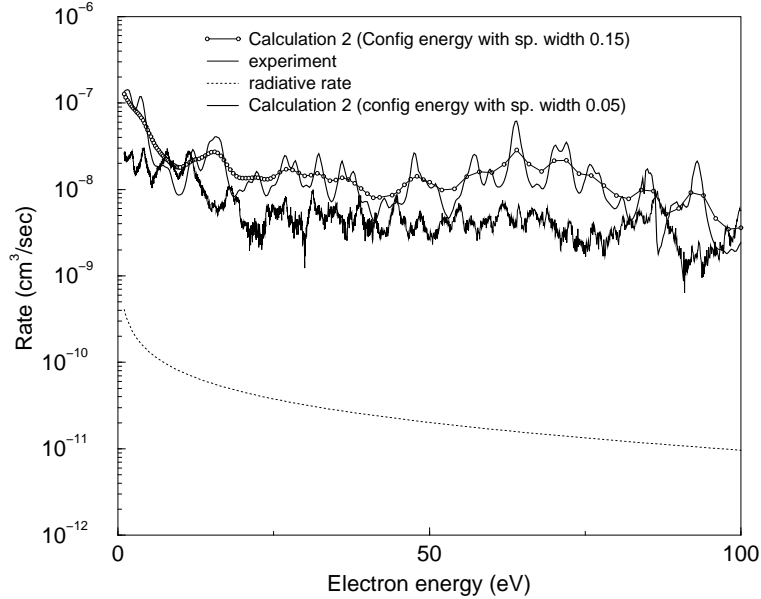}
\caption{Electron recombination rate on U$^{28+}$. Open circles
connected by solid line: Calculation 2 (using configuration energy and
$\Gamma_{spread}=.15$ a.u.), Solid line: Calculation 2 (using
configuration energy and $\Gamma_{spread}=0.05 $ a.u.), Dense solid line : Experiment, and Dotted line: Radiative rate.}
\end{figure}
To compare with Au$^{25+}$, we calculated the rate coefficients for the
electron energy ranging from 1 eV to 100 eV. 
\begin{figure}[h]
\epsfxsize=10cm
\centering\leavevmode\epsfbox{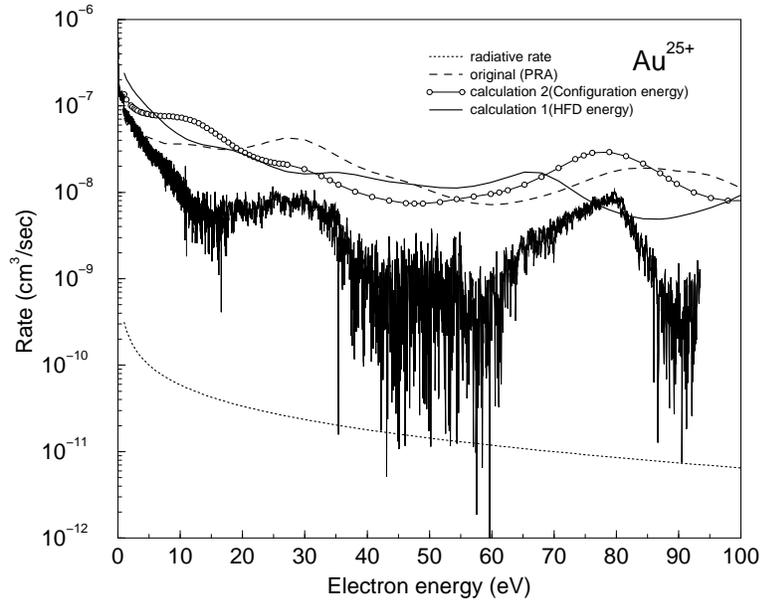}
\caption{Electron recombination rate on Au$^{25+}$. Open circles
connected by solid line: Calculation 2 (using configuration energy and
$\Gamma_{spread}=0.5$ a.u., Thick Solid line (using HFD energy and $\Gamma_{spread}=0.5$ a.u.), Thick dotted line: Original calculation (as in PRA {\bf 66}, 012713 (2002)), Dense solid line : Experiment, and Dotted line : Radiative rate.}.
\end{figure}
Because the previous
calculation\cite{Fl:02} was restricted to the energy range below 1
eV. Figure 9 shows results three different calculations and the
experimental data. It may be seen that the present (we call our best
calculation) calculation uses configuration energies shows a very good
qualitative agreement with the experiment in comparison to the other two
calculations. The RR is found be of smaller in magnitude. The most
striking feature in this graph is that the experimental data quickly
departs from the theoretical values as one goes to higher energies. This
can be interpreted as that in this energy range the fluorescence yield
certainly goes down from 1 which can be well understood from the Figure
10. In this figure we show the configuration energies as a function of
total angular momentum J. It shows that due to the presence of a lots
levels, many inelastic channels open up and they participate
strongly. This situation makes the fluorescence yield less than 1 and
hence the theory predicts recombination rates higher in magnitude. 

\begin{figure}[h]
\epsfxsize=10cm
\centering\leavevmode\epsfbox{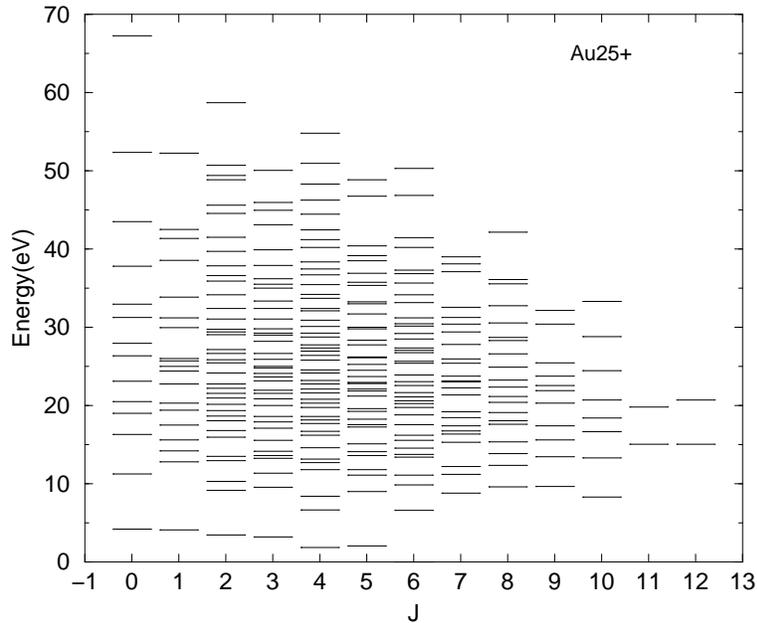}
\caption{Spectrum of low lying excited states of Au$^{25+}$. All
the levels shown belong to the 4f$^{8}$ configurations. The ground state is characterized by J=6.}  
\end{figure}

Which in turn just opposite to the case of U$^{28+}$ as shown in Figure 11.
\begin{figure}[h]
\epsfxsize=10cm
\centering\leavevmode\epsfbox{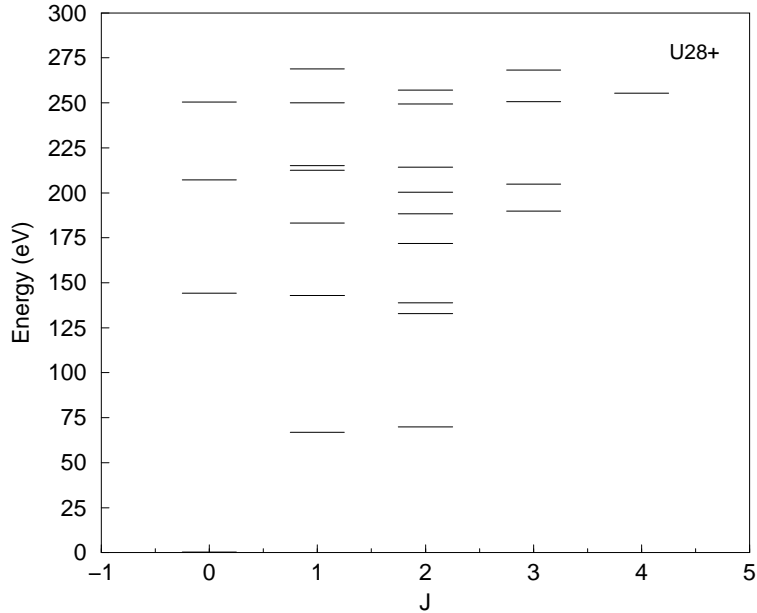}
\caption{Spectrum of low lying excited states of U$^{28+}$. All
the levels shown belong to 5s$^{2}$5p$^{2}$, 5s5p$^{3}$ and 5s$^{2}$5p5d
configurations (see table II). Its ground state is characterized by J=0.}
\end{figure}

Finally, we compare in table II, the present energies of the lowest configurations in U$^{28+}$ with the energies calculated by using HULLAC code\cite{Mitnik:98}. As mentioned earlier that in Ref\cite{Mitnik:98}, energies have been calculated by using different codes such as multiconfiguration Hartree-Fock (MCHF), the AUTOSTRUCTURE code in the perturbative-relativistic [AS(PR)] and in the semirelativistic [AS(SR)] mode and the HULLAC code. On comparison we found the present energies are close to those obtained by Mitnik et. al. \cite{Mitnik:98} using HULLAC code since both the calculations are fully relativistic.   
\begin{table}

\caption{Energies of the lowest configurations in U$^{28+}$. The present calculated energies obtained by using Hartree Fock Dirac (HFD) code are compared with the energies calculated using HULLAC code [18]. The energies are in eV.}
\label{tab:en_con}
\begin{tabular}{cccc}
Configurations & Term(J) &HULLAC & HFD (present)\\ 
\hline
$5s^{2}5p^{2}$ & 0&0.00 & 0.00\\
$5s^{2}5p^{2}$ & 1&65.884 & 66.980\\
$5s^{2}5p^{2}$ & 2&68.876 & 69.904\\
$5s5p^{3}$ & 2& 131.80 & 132.94\\
$5s^{2}5p^{2}$ & 2& 137.24 & 138.80\\
$5s5p^{3}$ & 1&142.05 & 142.97\\
$5s^{2}5p^{2}$ & 0& 143.60 & 144.16\\
$5s^{2}5p5d$& 2& 170.33 & 171.83\\
$5s^{2}5p5d$& 1&182.26 & 183.27\\
$5s^{2}5p5d$ & 2&186.37& 188.24\\
$5s^{2}5p5d$ & 3& 188.07 & 189.82\\
$5s5p^{3}$ & 2&198.49 & 200.24\\
$5s5p^{3}$ & 3&202.91 & 204.78\\
$5s5p^{3}$ & 0&205.43 & 207.29\\

$5s5p^{3}$ & 1&210.77 & 212.53\\

$5s5p^{3}$ & 2&212.50 & 214.21\\

$5s5p^{3}$ & 1&213.51 & 215.08\\

$5s^{2}5p5d$& 2&247.30 & 249.28\\

$5s5p^{3}$ & 1&287.25 & 250.11\\

$5s^{2}5p5d$ &3& 248.65 & 250.66\\

$5s^{2}5p5d$ & 0&248.40 & 250.40\\

$5s^{2}5p5d$ & 4& 252.64 & 255.34\\
$5s^{2}5p5d$ & 2& 254.29 & 257.00\\
$5s^{2}5p5d$ & 3& 266.20 & 268.30\\
$5s^{2}5p5d$ & 1& 266.67 & 268.86\\
\end{tabular}
\end{table}    
\section{Summary and outlook}
We have shown that the dielectronic states weakly mix with each other
and show a substantial mixing with complicated multiply excited
states. This explains the mechanism of low energy recombination of
electron with U$^{28+}$. The present results are found to be in good
agreement with the experimental data. We found that one must not ignore
the excitations from 4$f$ orbitals while considering the low energy
recombination of U$^{28+}$ ion. On detailed study, we
found the configuration mixing between the doubly excited states and
multiply excited states is not complete or uniform. Although this theory
is valid in a system having an extreme degree of configuration mixing
\cite{Fl:02}, still it predicts quite good results for other systems
where there is not enough strong configuration mixing, for example U$^{28+}$. This work develops further a statistical theory to study the low energy recombination process.  
\section{acknowledgements}
We thank Prof. A. M\"uller for providing experimental data of Au$^{25+}$
and U$^{28+}$ in numerical form and for his stimulating discussions. We also
thank to Dr. C. Harabati for allowing us to make use of his codes as
well as for his useful comments. Financial help from EPSRC is highly
acknowledged.


\begin{thebibliography}{99}
\bibitem{comment3}This fit is valid for $E>1$ a.u. In the Fermi-gas model
$a$ is related to the single-particle level density at the Fermi level
$g(\varepsilon _F)=3a^2/2\pi^2$. The fitted value $a=3.99$ gives
$g(\varepsilon _F)=2.4$ a.u., in agreement with the orbital spectrum in
Fig.1
\bibitem{comment4} Here we assume that the electron energy is below the target
excitation threshold.
\end{thebibliography}
\end{document}